\documentclass[twocolumn,showpacs,preprintnumbers,amsmath,aps]{revtex4}
\usepackage{graphicx}
\usepackage{dcolumn}
\usepackage{bm}
\usepackage{color}
\begin{document}
%
%
\def\eq#1{(\ref{#1})}
\def\fig#1{Fig.\hspace{1mm}\ref{#1}}
\def\tab#1{table\hspace{1mm}\ref{#1}}
%
\title{The high-pressure superconductivity in SiH$_4$: the strong-coupling approach}
\author{R. Szcz{\c{e}}{\'s}niak, A.P. Durajski}
\affiliation{Institute of Physics, Cz{\c{e}}stochowa University of Technology, Ave. Armii Krajowej 19, 42-200 Cz{\c{e}}stochowa, Poland}
\date{\today} 
\begin{abstract}
In the paper, the thermodynamic parameters of the high-pressure superconducting state in the SiH$_4$ compound have been determined ($p=250$ GPa). By using the Eliashberg equations in the mixed representation, the critical temperature, the energy gap, and the electron effective mass have been calculated. It has been stated that the critical temperature ($T_{C}$) decreases from $51.65$ K to $20.62$ K, if the Coulomb pseudopotential increases ($\mu^{\star}\in\left<0.1,0.3\right>$). The dimensionless ratio $2\Delta\left(0\right)/k_{B}T_{C}$ decreases from $4.10$ to $3.84$, where the symbol $\Delta\left(0\right)$ denotes the value of the order parameter close to the zero temperature. The ratio of the electron effective mass to the band electron mass is high, and it reaches maximum equal to $1.95$ for the critical temperature.
\end{abstract}
\maketitle

The metallic state in the hydrogen is intensely studied since 1935 \cite{Wigner}. It has been shown that the hydrogen is metallic under the high compression ($\sim 400$ GPa) \cite{Stadele}. 

At the low value of the temperature, the metallic state of the hydrogen transforms into the superconducting state \cite{Ashcroft}. In the pressure ($p$) range from $\sim 400$ GPa to $\sim 500$ GPa, the value of the critical temperature ($T_C$) reaches maximum equal to $242$ K  ($p=450$ GPa) \cite{Cudazzo,SzczesniakActa2012}. For the pressure's values from $\sim 500$ GPa to $\sim 800$ GPa, the critical temperature increases: $T_{C}\in\left<282, 360\right>$ K \cite{Yan,SzczesniakDominH}. The extremely high value of the critical temperature has been predicted near $2000$ GPa, where $T_{C}$ can assumes even 
$\sim 600$ K \cite{Maksimov,Szczesniak-H2000}.

Recently, it has been suggested that the hydrogen-rich compounds become metallic at the pressure's value lower than for the pour hydrogen. In particular, the following compounds have been taken under consideration: methane (CH$_4$) \cite{Lin}, silane (SiH$_4$, SiH$_4$(H$_2$)$_2$) \cite{Chen,Eremets,Li}, disilane (Si$_2$H$_6$) \cite{Jin}, and germane (GeH$_4$, GeH$_4$(H$_2$)$_2$) \cite{Zhang,Zhong}.  

In the case of SiH$_4$(H$_2$)$_2$ under pressure at $270$ GPa and Si$_2$H$_6$ at $p=250$ GPa, the theoretical results have predicted the existence of the high-temperature superconducting state: $T_{C}=129.83$ K and $T{_C}=173.36$ K, respectively \cite{DurajskiSiH8}, \cite{DurajskiSiH6}. We notice that the above values of the critical temperature are even higher than for the cuprates \cite{SzczesniakPlos}.

In the presented paper, we have studied the thermodynamic properties of the superconducting state in the SiH$_4$ compound under the pressure at $250$ GPa.

We notice that the experimental data have confirmed the metallization in SiH$_4$ ($p\simeq 50$ GPa) \cite{Chen,Eremets}. Furthermore, the critical temperature increases with the pressure and assumes the maximum equal to $17.5$ K at $96$ GPa and at $120$ GPa. 

In the SiH$_4$ compound ($p=250$ GPa), the strong electron-phonon interaction has been predicted \cite{Chen2008}. In particular, the electron-phonon coupling constant takes the value: $\lambda=0.91$. Taking into consideration, the above result one should expect the high value of the critical temperature. In the considered case, the remaining thermodynamic parameters are probably beyond the BCS predictions \cite{BCS1}, \cite{BCS2}. For this reason, the numerical calculations have been made in the framework of the Eliashberg approach \cite{Eliashberg}. 

The Eliashberg equations defined both on the real and imaginary axis (the mixed representation) can be written in the following form \cite{Marsiglio}, \cite{Daams}:
\begin{eqnarray}
\label{r1}
\phi\left(\omega\right)&=&
\frac{\pi}{\beta}\sum_{m=-M}^{M}\frac{\left[\lambda\left(\omega-i\omega_{m}\right)-\mu^{\star}\left(\omega_{m}\right)\right]}
{\sqrt{\omega_m^2Z^{2}_{m}+\phi^{2}_{m}}}\phi_{m}\\ \nonumber
                              &+& i\pi\int_{0}^{+\infty}d\omega^{'}\alpha^{2}F\left(\omega^{'}\right)
                                  [\left[N\left(\omega^{'}\right)+f\left(\omega^{'}-\omega\right)\right]\\ \nonumber
                              &\times&K\left(\omega,-\omega^{'}\right)\phi\left(\omega-\omega^{'}\right)]\\ \nonumber
                              &+& i\pi\int_{0}^{+\infty}d\omega^{'}\alpha^{2}F\left(\omega^{'}\right)
                                  [\left[N\left(\omega^{'}\right)+f\left(\omega^{'}+\omega\right)\right]\\ \nonumber
                              &\times&K\left(\omega,\omega^{'}\right)\phi\left(\omega+\omega^{'}\right)],
\end{eqnarray}
and
\begin{eqnarray}
\label{r2}
Z\left(\omega\right)&=&
                                  1+\frac{i\pi}{\omega\beta}\sum_{m=-M}^{M}
                                  \frac{\lambda\left(\omega-i\omega_{m}\right)\omega_{m}}{\sqrt{\omega_m^2Z^{2}_{m}+\phi^{2}_{m}}}Z_{m}\\ \nonumber
                              &+&\frac{i\pi}{\omega}\int_{0}^{+\infty}d\omega^{'}\alpha^{2}F\left(\omega^{'}\right)
                                  [\left[N\left(\omega^{'}\right)+f\left(\omega^{'}-\omega\right)\right]\\ \nonumber
                              &\times&K\left(\omega,-\omega^{'}\right)\left(\omega-\omega^{'}\right)Z\left(\omega-\omega^{'}\right)]\\ \nonumber
                              &+&\frac{i\pi}{\omega}\int_{0}^{+\infty}d\omega^{'}\alpha^{2}F\left(\omega^{'}\right)
                                  [\left[N\left(\omega^{'}\right)+f\left(\omega^{'}+\omega\right)\right]\\ \nonumber
                              &\times&K\left(\omega,\omega^{'}\right)\left(\omega+\omega^{'}\right)Z\left(\omega+\omega^{'}\right)], 
\end{eqnarray}
%

%
\begin{figure*}[th]
\includegraphics[scale=0.45]{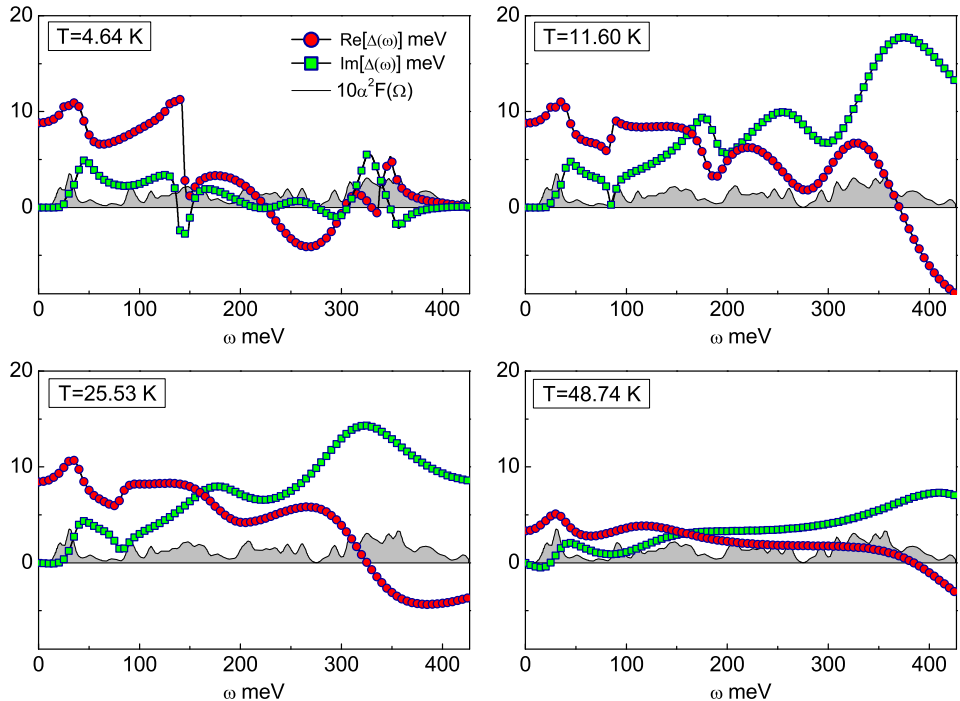}
\caption{The real and imaginary part of the order parameter for selected values of the temperature and $\mu^{\star}=0.1$. In the figure, the rescaled Eliashberg function has been also plotted.}
\label{f1}
\end{figure*}
%

where:
\begin{equation}
\label{r3}
K\left(\omega,\omega^{'}\right)\equiv
\frac{1}{\sqrt{\left(\omega+\omega^{'}\right)^{2}Z^{2}\left(\omega+\omega^{'}\right)-\phi^{2}\left(\omega+\omega^{'}\right)}}.
\end{equation}

The symbols $\phi\left(\omega\right)$ and $Z\left(\omega\right)$ denote the order parameter function and the wave function renormalization factor on the real axis ($\omega$); $\phi_{m}\equiv\phi\left(i\omega_{m}\right)$ and $Z_{m}\equiv Z\left(i\omega_{m}\right)$ represent the values of these functions on the imaginary axis ($i\equiv\sqrt{-1}$). The Matsubara frequency is given by: $\omega_{m}\equiv \left(\pi / \beta\right)\left(2m-1\right)$, where $\beta\equiv\left(k_{B}T\right)^{-1}$ ($k_{B}$ is the Boltzmann constant). Both on the real and imaginary axis, the order parameter is defined as: $\Delta\equiv \phi/Z$. 

The electron-phonon pairing kernel has the form: 
\begin{equation}
\label{r4}
\lambda\left(z\right)\equiv 2\int_0^{\Omega_{\rm{max}}}d\Omega\frac{\Omega}{\Omega ^2-z^{2}}\alpha^{2}F\left(\Omega\right),
\end{equation}
where the Eliashberg function ($\alpha^{2}F\left(\Omega\right)$) has been calculated in the paper \cite{Chen2008}; the maximum phonon frequency ($\Omega_{\rm{max}}$) is equal to $426.1$ meV.

The function $\mu^{\star}\left(\omega_{m}\right)\equiv\mu^{\star}\theta\left(\omega_{c}-|\omega_{m}|\right)$ describes the electron depairing interaction; 
$\mu^{\star}$ is the Coulomb pseudopotential. Due to the absence of the experimental value of the critical temperature, the Coulomb pseudopotential is unknown. For this reason, we have assumed: $\mu^{\star}\in\left<0.1,0.3\right>$. The symbol $\theta$ denotes the Heaviside unit function and $\omega_{c}$ is the cut-off frequency ($\omega_{c}=3\Omega_{\rm{max}}$).
 
The quantities N($\omega$) and f($\omega$) denote the Bose-Einstein and Fermi-Dirac function, respectively.

The order parameter function and the wave function renormalization factor on the imaginary axis have been calculated by using the equations \cite{Eliashberg}:
\begin{equation}
\label{r5}
\phi_{n}=\frac{\pi}{\beta}\sum_{m=-M}^{M}
\frac{\lambda\left(i\omega_{n}-i\omega_{m}\right)-\mu^{\star}\left(\omega_{m}\right)}
{\sqrt{\omega_m^2Z^{2}_{m}+\phi^{2}_{m}}}\phi_{m},
\end{equation}
\begin{equation}
\label{r6}
Z_{n}=1+\frac{1}{\omega_{n}}\frac{\pi}{\beta}\sum_{m=-M}^{M}
\frac{\lambda\left(i\omega_{n}-i\omega_{m}\right)}{\sqrt{\omega_m^2Z^{2}_{m}+\phi^{2}_{m}}}\omega_{m}Z_{m}.
\end{equation}

The Eliashberg set has been solved for $M=1100$. We have used the numerical methods presented in the papers \cite{Szczesniak01} and \cite{Szczesniak02}.
In the considered case, the solutions of the Eliashberg equations are stable for $T\geq T_{0}=4.64$ K. 

In \fig{f1}, the form of the order parameter on the real axis has been presented for the selected values of the temperature and $\mu^{\star}=0.1$.
Moreover, the rescaled Eliashberg function (10$\alpha^{2}$F($\Omega$)) has been also plotted. 

It is easy to see that for the low frequencies, the non-zero values are taken only by the real part of the order parameter. The obtained result indicates that in the considered range of frequencies the damping effects related with Im$[\Delta(\omega)]$ not exist. Additionally, it has been stated that the shapes of the functions Re$[\Delta(\omega)]$ and Im$[\Delta(\omega)]$ are correlated with the complicated form of the Eliashberg function. 

The physical value of the order parameter has been calculated by using the equation:
\begin{equation}
\label{r7}
\Delta\left(T\right)={\rm Re}\left[\Delta\left(\omega=\Delta\left(T\right)\right)\right].
\end{equation}

\begin{figure}[ht]
\includegraphics*[width=\columnwidth]{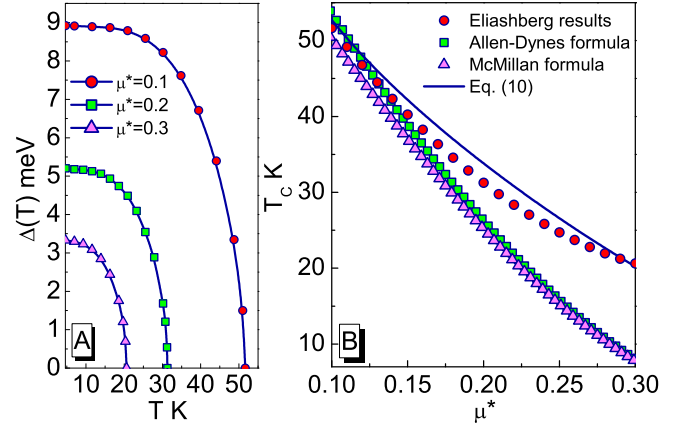}
\caption{
(A) The dependence of the order parameter on the temperature for selected values of the Coulomb pseudopotential.
(B) The critical temperature as a function of the Coulomb pseudopotential. 
The circles represent the results obtained by using the Eliashberg equations. 
The squares and triangles are related to the classical Allen-Dynes and McMillan expression. 
The solid line has been achieved with help of Eq. (10). 
}
\label{f2}
\end{figure}

In \fig{f2} (A), the dependence of the order parameter on the temperature for the selected values of the Coulomb pseudopotential has been presented. We notice that the value $\Delta\left(T,\right)$ can be parameterized by using the formula:
\begin{equation}
\label{r8}
\Delta\left(T\right)=\Delta\left(T_{0}\right)\sqrt{1-\left(\frac{T}{T_{C}}\right)^{\beta}},
\end{equation}
where $\beta=3.4$, and:
\begin{equation}
\label{r9}
\Delta\left(T_{0}\right)=91.93\left(\mu^{\star}\right)^{2}-64.73\mu^{\star}+14.47 \quad {\rm meV}.
\end{equation}

Next, we have determined the dependence of the critical temperature on the Coulomb pseudopotential. The results have been presented in \fig{f2} (B). 

It is easy to notice that the value of $T_{C}$ is high in the whole range of the considered values of $\mu^{\star}$. In particular, $T_{C}\in\left<51.65,20.62\right>$ K. 

Additionally, we underline that for large value of the Coulomb pseudopotential, the critical temperature can not be precisely estimated by the classical Allen-Dynes or McMillan formula \cite{AllenDynes}, \cite{McMillan}. However, the modified Allen-Dynes formula derived originally for ${\rm SiH_4(H_2)_2}$ compound works very well \cite{DurajskiSiH8}:
\begin{equation}
\label{r10}
k_{B}T_{C}=f_{1}f_{2}\frac{\omega_{{\rm ln}}}{1.37}\exp\left[\frac{-1.125\left(1+\lambda\right)}{\lambda-\mu^{\star}}\right],
\end{equation}
where $f_{1}$ and $f_{2}$ denote the functions \cite{AllenDynes}: 
\begin{equation}
\label{r11}
f_{1}\equiv\left[1+\left(\frac{\lambda}{\Lambda_{1}}\right)^{\frac{3}{2}}\right]^{\frac{1}{3}}, \quad {\rm and} \quad
f_{2}\equiv 1+\frac{\left(\frac{\sqrt{\omega_{2}}}{\omega_{\rm{ln}}}-1\right)\lambda^{2}}{\lambda^{2}+\Lambda^{2}_{2}}.
\end{equation}

The quantities $\Lambda_{1}$ and $\Lambda_{2}$ take the form:
$\Lambda_{1}=2-0.14\mu^{\star}$, and $\Lambda_{2}=\left(0.27+10\mu^{\star}\right)\left(\sqrt{\omega_{2}}/\omega_{\ln}\right)$. 

The second moment of the normalized weight function and the logarithmic phonon frequency are given by: 
\begin{equation}
\label{r12}
\omega_{2}\equiv\frac{2}{\lambda}\int^{\Omega_{\rm{max}}}_{0}d\Omega\alpha^{2}F\left(\Omega\right)\Omega,
\end{equation}
and
\begin{equation}
\label{r13}
\omega_{{\rm ln}}\equiv \exp\left[\frac{2}{\lambda}
\int^{\Omega_{\rm{max}}}_{0}d\Omega\frac{\alpha^{2}F\left(\Omega\right)}
{\Omega}\ln\left(\Omega\right)\right].
\end{equation}

The electron-phonon coupling constant has the form:
\begin{equation}
\label{r14}
\lambda\equiv 2\int^{\Omega_{\rm{max}}}_{0}d\Omega\frac{\alpha^{2}F\left(\Omega\right)}{\Omega}.
\end{equation}
For ${\rm SiH_{4}}$ under the pressure at $250$ GPa, it has been achieved: $\sqrt{\omega_{2}}=167.64$ meV and $\omega_{{\rm ln}}=72.42$ meV.

%
%

%
\begin{figure}[ht]
\includegraphics*[width=\columnwidth]{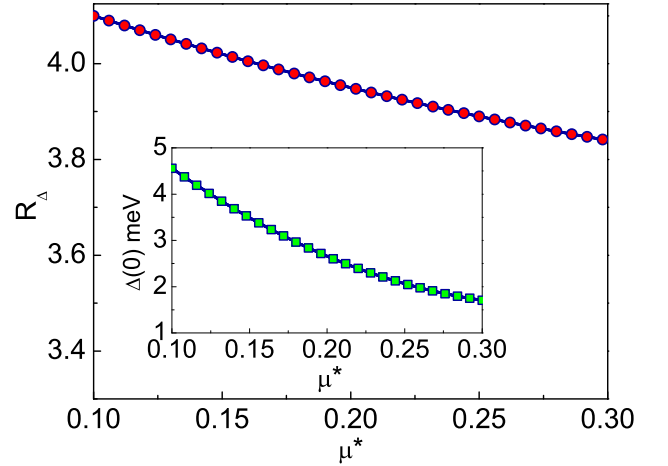}
\caption{The dependence of the ratio $R_{\Delta}$ on the value of the Coulomb pseudopotential. In the inset, the influence of $\mu^{\star}$ on $\Delta(0)$ has been presented.}
\label{f3}
\end{figure}
%

Now, we consider the low-temperature value of the order parameter at the Fermi level ($\Delta\left(0\right)\equiv\Delta\left(T_{0}\right)$). In particular, we have $\Delta\left(0\right)\in\left<4.56,1.71\right>$ meV for $\mu^{\star}\in\left<0.1,0.3\right>$. The full dependence of $\Delta\left(0\right)$ on the Coulomb pseudopotential has been shown in the inset in \fig{f3}.

Taking into consideration the obtained results, we have calculated the dimensionless ratio 
$R_{\Delta}\equiv 2\Delta\left(0\right)/k_{B}T_{C}$. It has been achieved: $R_{\Delta}\in\left<4.10,3.84\right>$ (see also \fig{f3}). 
We notice that the values of $R_{\Delta}$ for SiH$_4$ compound differ significantly from the prediction of the BCS model, where $R_{\Delta}=3.53$ \cite{BCS1,BCS2}. 

%
\begin{figure}[ht]
\includegraphics*[width=\columnwidth]{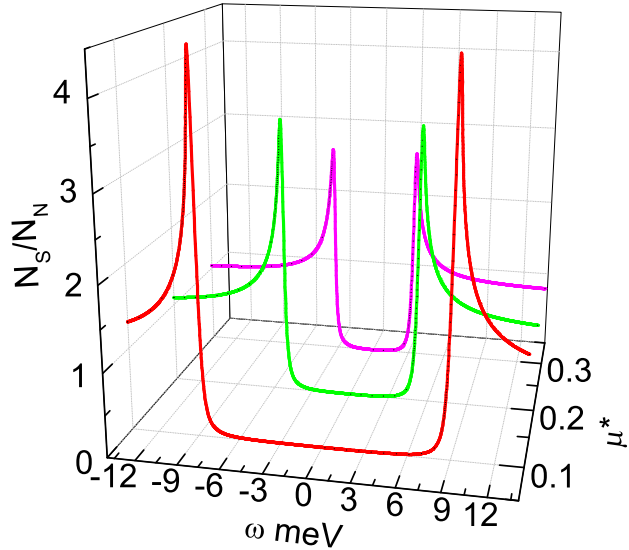}
\caption{The dependence of the total normalized density of states on the frequency for $T=T_{0}$. The selected values of the Coulomb pseudopotential have been assumed.}
\label{f4}
\end{figure}
%

%
\begin{figure}[ht]
\includegraphics*[width=\columnwidth]{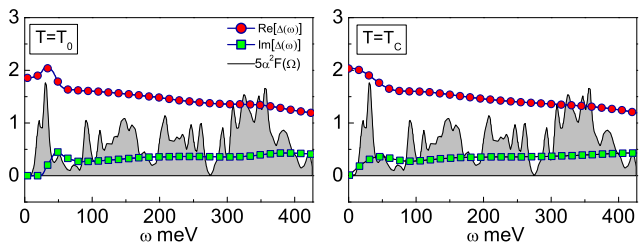}
\caption{
The dependence of the real and imaginary part of the wave function renormalization factor on the frequency for $T=T_{0}$ and $T=T_{C}$. 
In the figure, the rescaled Eliashberg function has been also plotted (5$\alpha^{2}F\left(\Omega\right)$).
The results have been obtained for $\mu^{\star}=0.1$.
}
\label{f5}
\end{figure}
%

The order parameter function on the real axis allows to calculate the total normalized density of states \cite{Eliashberg}:
\begin{equation}
\label{r15}
\frac{N_{S}\left(\omega \right)}{N_{N}\left(\omega \right)}={\rm Re}\left[\frac{\left|\omega -i\Gamma \right|}{\sqrt{\left(\omega -i\Gamma\right)^{2}}-\Delta^{2}\left(\omega\right)}\right],
\end{equation}
where the symbols $N_{S}$ and $N_{N}$ denote the density in the superconducting and normal state, respectively. The pair breaking parameter $\Gamma$ is equal to $0.15$ meV. 

The plot of the total normalized density of states has been presented in \fig{f4}. We have assumed $T=T_{0}$ and the selected values of $\mu^{\star}$.
It is easy to see the reduction of the function $N_{S}\left(\omega \right)/N_{N}\left(\omega \right)$ by the depairing electron correlations.

In \fig{f5}, the real and imaginary part of the wave function renormalization factor on the real axis has been presented ($T=T_{0}$ and $T=T_{C}$).

On the basis of the obtained results, it has been stated that $Z\left(\omega\right)$ weakly depends on the temperature in comparison with the order parameter. We can also observe the weak correlation between the wave function renormalization factor and the shape of the Eliashberg function.

The dependence of the electron effective mass ($m^{\star}_{e}$) on the temperature has been determined by using the expression: 
$m^{\star}_{e}={\rm Re}\left[Z\left(0\right)\right]m_{e}$, where $m_{e}$ denotes the band electron mass.

In our case, the results prove that the electron effective mass takes the high value in the whole range of the superconducting phase's existence. The maximum of $m^{\star}_{e}$ has been observed for $T=T_{C}$, where $m^{\star}_{e}=1.95m_{e}$. We notice that the Coulomb pseudopotential does not influence on the value of $\left[m^{\star}_{e}\right]_{\rm max}$. 

\vspace*{0.25cm}

In the paper, the Eliashberg equations for the SiH$_4$ compound under the pressure at $250$ GPa have been solved.
We have shown that the value of the critical temperature decreases from 51.65 K to 20.62 K if $\mu^{\star}\in\left<0.1,0.3\right>$.
The dimensionless ratio $R_{\Delta}$ takes the values beyond the prediction of the BCS model: $R_{\Delta}\in\left<4.10,3.84\right>$.
Additionally, it has been stated that the electron effective mass takes the high values and $\left[m^{\star}_{e}\right]_{\rm max}$ is equal to $1.95m_{e}$ for the critical temperature.

\vspace*{0.25cm}

{\bf Acknowledgments}

The authors wish to thank Prof. K. Dzili{\'n}ski for providing excellent working conditions and the financial support.

All numerical calculations have been based on the Eliashberg function sent to us by Prof. Xiao-Jia Chen for whom we are very thankful.

The authors are grateful to the Czestochowa University of Technology - MSK CzestMAN for granting access to the computing infrastructure built in the project No. POIG.02.03.00-00-028/08 "PLATON - Science Services Platform".

%

\end{document}